\documentclass{article}
\usepackage{graphicx} % Required for inserting images
\usepackage{float}

\usepackage{comment}
\usepackage[sorting=none,
backend=biber,
hyperref=false,
backref=false]{biblatex}          % packages for bibliography
\DeclareUnicodeCharacter{0308}{XXX HERE I AM XXX ???}

\usepackage{caption}
\usepackage{subcaption}
\usepackage{comment}
\usepackage{multirow}

\title{Determining social mechanisms for sequential decision-making in a virtual pedestrian route choice experiment}
\author{ Anna Sigalou\textsuperscript{1,3}\footnote{Corresponding author: anna.sigalou@ipla.csic.es}, Yunhe Tong\textsuperscript{4,2}\footnote{Corresponding author: yunhe.tong@bristol.ac.uk}, Charlie Pilgrim\textsuperscript{1},\\ Richard P. Mann\textsuperscript{1}, Nikolai W.F. Bode\textsuperscript{2}\footnote{Corresponding author: nikolai.bode@bristol.ac.uk}}
\date{ \small
\textsuperscript{1}School of Mathematics, University of Leeds, UK\\
\textsuperscript{2}School of Engineering Mathematics and Technology, University of Bristol, UK\\
\textsuperscript{3}Instituto de Productos Lácteos de Asturias, CSIC, Spain\\
\textsuperscript{4}School of Public Affairs, University of Science and Technology of China, Hefei 230026, PR China\\}

\begin{document}

\maketitle

%\newpage

\begin{abstract}
    Moving groups are routinely faced with a choice of different routes as part of their daily lives, such as choosing between exits from a building.  Differences in moving speeds and environmental constraints often lead to individuals being able to observe the choices of others ahead. This social information can inform their decision-making, but exactly how this is being used remains an open question. Previous theoretical studies on animal groups have demonstrated that simple heuristics are plausible and accurate mechanisms, with some further predicting that more recent decisions are more heavily weighted. Experiments with fish corroborate the importance of more recent decisions; however, experimental work is limited. Here, we conduct an online survey with human participants to identify which social decision-making mechanism individuals follow. Contrary to previous experimental work, we find little indication that recent decisions are weighted more heavily; instead, our results suggest that following the majority of previous decisions is the dominant behaviour. Furthermore, self-reported decision-making mechanisms correlate with our experimental findings despite their variability, suggesting that on average individuals can recognize their behavioural tendencies. Our findings give insight into social sequential decision-making, and provide an empirical foundation for integrating realistic social decision mechanisms into pedestrian movement models.
    
\end{abstract}


\section{Introduction}

Picture a crowd of people leaving a building. As they walk along a corridor, the crowd stretches out in length, causing individuals to walk behind each other. At the end of the corridor are two exits next to each other. One after the other, individuals choose exits and leave the building. This everyday scenario occurs across contexts, including students leaving teaching buildings, passengers walking through transport hubs or metro stations, or workers evacuating buildings during fire alarms. It is also common in moving animal groups, such as schools of fish moving in shallow-water channels or ungulates moving around obstacles. What is common to all contexts is that individuals have to decide on a route, and in addition to their own beliefs, observations, or preferences they also consider the choices made by others moving in front of them. This results in a sequence of decisions, where the role of social information can be crucial to determine the distribution of individuals across routes. For example, if everyone follows the first person, one route remains unused, and if no-one follows others, equivalent routes are used equally.

Such social mechanisms can be beneficial with individuals in groups making better decisions than they would on their own, as others can provide valuable information about the environment leading to fewer mistakes and increased accuracy \cite{wolf2013accurate}. They may also be detrimental, for example, if environmental information is very limited as this may lead to erroneous cascades \cite{Giraldeau2002} or when routes are over-used, as can happen in emergency evacuations \cite{fahy2011station}. The collective decision-making depends on how individuals balance their personal information with the social information that is being provided through the group; exactly how this is done is still debated.

Considering the sequential decisions we investigate here, theoretical work suggests a unified modelling framework that can incorporate different decision-making mechanisms that refer to the way available social information is perceived or used \cite{Perez-Escudero2011, Arganda2012}. It incorporates the main principles of personal information, social information, and social behaviour. While social information is described as an ordered sequence, some mechanisms consider unordered simplifications, such as the difference in decisions between two options thus far. Further theoretical development suggests that sequence-unaware heuristic approaches, where individuals are unaware of the ordering of the previous decisions, can approximate the results obtained by sequence-aware mechanisms \cite{perez2013estimation}. However, it has also been suggested that more recent decisions may be weighed more heavily \cite{Mann2018, mann2021}. Theoretical work has thus introduced a range of possible mechanisms and models that need to be tested on data.

Experimental work mirrors this variety. Static information, such as the number of individuals already on an option, can lead to a linear increase in the probability that a focal individual will follow \cite{ward2013initiators}, but there is also evidence that more recently observed decisions are more important; for example, work on humbug damselfish suggests that they respond to `dynamic' information, such as recent movements from their neighbours, rather than `static' information, such as the current positions of these neighbours at a specific time \cite{mann2014humbugs}. However, there is very limited research on a strictly sequential decision-making setting that explores this question, with the experiments of Kadak and Miller with zebrafish \cite{Kadak2020} being the only one so far (to the best of our knowledge). Their findings corroborate the notion that dynamic information is followed, by suggesting that zebrafish following an ordered sequence of prior decisions seem to rely on the most recent decision \cite{Kadak2020}. Both Mann \cite{Mann2018} and Kadak and Miller \cite{Kadak2020} point out that we do not know what the internal processes of the animals they studied are. As such, it is possible that the mechanisms identified may simply be a good match to their data without describing the underlying cognitive processes or more general behaviour of animals. Here, we seek to contribute to this body of work in two ways. First, we contribute experimental data from a different species (humans), and second, making use of humans' ability to introspect on the reasons for their behaviour, we can contrast our experimental findings to the self-reported decision mechanisms by participants.

The scenario we consider in this work is that of a group of people evacuating a building during an emergency. This is one example of pedestrian route choice, a key aspect of pedestrian crowd dynamics \cite{tong2022principles}. Social mechanisms in pedestrian route choice determine the use of pedestrian facilities, including the distribution of evacuees over exits during emergencies. Understanding them is thus crucial for the management and planning of pedestrian facilities. Previous work has investigated the type of information used in pedestrian route choice \cite{basu2022systematic}. Theoretical and experimental work considers the influence on route choice of many different factors that describe differences between routes, relating to expected travel times, crowd density, distances, signage, and the familiarity of individuals with routes (e.g. \cite{gabbana2022fluctuations, tong2025exploring, li2019comparing, kinateder2018exit, ronchi2016variable}). To give an example, it has been suggested that pedestrians follow dynamic rather than static information, where aspects, such as the speed and length of queues at exits, may be more important than their width \cite{bode2015information}. However, previous research on exit choice during evacuations often treats social information as part of the overall properties of a route, typically measured through crowd-related indicators such as crowd size \cite{kinateder2021exit}, perceived busyness \cite{klamroth2020network} and the proportion of people selecting a given exit \cite{lin2020people,tong2021value}. Such studies tend to depend strongly on the context \cite{haghani2017following,}. For example, when social information (e.g., the movement direction of a crowd) serves as the primary signal for participants who lack other cues, individuals are more likely to follow the crowd \cite{moussaid2016crowd}. In contrast, when participants already know which route leads to safety, they may avoid crowded exits to enhance their evacuation efficiency \cite{lovreglio2016study}. Nevertheless, the focus has been on the properties of different routes, rather than on an observed sequence of decisions for different options. As such, our work addresses this gap by investigating how individuals respond to sequences of social information during evacuation scenarios.

To examine this, we use a virtual experiment for our data collection. Virtual experiments are a well-accepted experimental paradigm in research on pedestrian route choice, because of their benefits in terms of ease of data collection, safety, and controllability. A range of different virtual experiment types are used \cite{feng2021different}, ranging from fully immersive virtual reality (e.g., where participants navigate the environment using VR headsets \cite{lovreglio2022exit}), to highly abstracted on-screen displays (e.g.,in which participants control a virtual avatar using a keyboard and/or mouse to perform navigation tasks on a computer screen \cite{zhao2018networked}). Previous research suggests that pedestrians’ exit choice behaviours remain consistent across various virtual experiment types \cite{ruddle2004effects,li2019comparing}.  In this work, we adopt a desktop-based setup, which provides sufficient ecological validity for capturing simple route choice behaviours while maintaining practical advantages such as ease of implementation, reduced participant fatigue, and minimal technical requirements. Moreover, previous work comparing behaviour in a virtual experiment and in real-world observations suggests this type of experiment is adequate for capturing simple route choices \cite{li2019comparing,vanbeek2024comparison}. Nevertheless, the extent to which findings from virtual experiments extend to real-world evacuations remains an open question \cite{kinateder2014virtual} and our contribution should therefore be viewed as a starting point for further investigation.

In this work, we aim to identify whether the dynamic heuristic rule of following the most recent decision is employed, and whether it is significantly favoured over some other rule, such as following the overall majority. By working with people, we gain insight into the reasoning behind decision-making which is not possible with animals. By keeping the scenario simple and focusing on a setting of sequential decisions, we can compare our findings with existing research, and we address a gap in the literature on pedestrian route choice. The novelty of this work lies in providing the human experimental test of sequential social decision-making within an evacuation context. Methodologically, it bridges theoretical models developed in animal and collective behaviour research with human crowd dynamics, offering a framework for analysing how individuals integrate sequential social cues during evacuation. The findings have potential applications for improving behavioural realism in evacuation modelling, refining simulation algorithms, and informing the design of communication strategies that account for dynamic social influence during emergencies.

In the following, we present our methods, starting with an introduction to sequential decision-making models, including key mechanisms for responding to sequential social information, before describing our experiment, data collection, and findings. 

\section{Methods}

\subsection{Sequential decision-making models}
We consider a sequential binary decision setting. The individuals within that setting have the same personal information $a$, as well as a tendency $s$ to follow the available social information \textit{I} provided by the actions of the previous decision-makers. Using that information, they make an estimate of which is the best out of the two available choices. The decision-making process performed by the individual, when option A is best, is described as the probability of choosing option A as such:

\begin{equation}
    P_A = \frac{1}{1+as^{-I}}
    \label{model_basic}
\end{equation}

The parameter $a$ represents the non-social information and in its original formulation defines the ratio between choosing option B over choosing option A when A is correct, given the non-social (personal) information \cite{Perez-Escudero2011}; it takes values $a > 0$ with reliability decreasing as $a$ increases. The value of $s$ describes how strongly social information is being followed and it takes values of $s > 1$ for active social following (with higher values of $s$ representing a higher intensity). The value of $I$ is determined by the amount of social information available to the focal decision-maker, and it depends on the decision-making strategy. This process is symmetric, with the probability of choosing the alternative, $B$, being defined as $P_B=1-P_A$.

This decision-making model was originally described and analysed by \cite{Perez-Escudero2011} and it is now widely used \cite{Arganda2012, sigalou2023evolutionary}. Fundamentally, social information is available as an ordered sequence of past decisions, but simpler decision-making strategies can make use of lower-dimensional representations such as the aggregate number of individuals choosing each option \cite{sigalou2023evolutionary}. Here we investigate three possibilities: aggregate strategy (number choosing each option); majority strategy (binary indicator of which option the majority has chosen), and dynamic strategy (only the most recent decision). A further strategy that makes use of the full sequential information has been derived by \cite{Perez-Escudero2011}; this derivation accounts for the full statistical dependency between past decisions, so we label it the dependencies strategy. We do not include this strategy in our empirical analysis as we regard it as implausible that the observer will memorise a full sequence of past choices and fully account for the dependencies between them in her own considerations; we note, however, that derivations of optimal decision-making when given a full sequence of past decisions indicate a strong dependence on the most recent decision \cite{Mann2018}, and this provides a theorectical motivation for considering the dynamic strategy \cite{mann2021, sigalou2023evolutionary} 

The aggregate strategy is the case where the focal agent observes the difference in numbers between the two options at the time of her decision; it was considered in the original publication of \cite{Perez-Escudero2011} together with the dependencies strategy. In this decision-making rule the focal agent observes the numbers of either option at the moment of her decision collapsing it from an ordered sequence to a single value, as it is now defined as $I=\Delta n$, where $\Delta n = n_A-n_B$ is the difference in agents that have chosen option $A$ minus the number of agents that have chosen option $B$. 

The majority strategy is the case where the focal agent observes the sign of the difference in numbers between the two options at the time of her decision. It is defined by $I=\textrm{sign}(\Delta n)$, where $\textrm{sign}(\Delta n)$ is the sign of $\Delta n$ defined above. This decision-making rule was considered by \cite{sigalou2023evolutionary} as a plausible simplified version of the aggregate strategy.

The dynamic strategy describes the case where the focal agent only considers the most recent decision and disregards older information, such that $I=d$, where $d = 1$ if the most recent choice was $A$ and $d=-1$ if $B$. This decision-making rule was considered by \cite{mann2021} as a way to capture the higher relevance that more recent decisions can have compared to outdated older decisions.

These decision-making strategies form hypotheses that can be compared against data to identify likely behaviours, and overlap with most of the strategies considered by \cite{Kadak2020}. We will do so for experimental data from human participants.

\subsection{Virtual Experiment}

To determine which decision rule is most likely employed when making a socially informed decision during sequential decision-making we conducted a virtual experiment with human participants. Before the experiment started, participants were provided with information about the experiment and about how their data would be handled, a consent form, and were asked to state their age and sex. Participants were told that ``You are evacuating a building during an emergency. You are moving in an orderly fashion, trying to find the exit. Wait while the people in front of you make their path decisions, and then quickly make yours!". This phrasing was chosen to gently but implicitly encourage the participants to try and follow the group rather than avoid it (eg to avoid congestion). They were then shown a video showing a route choice scenario with computer-generated pedestrians and after its completion they were asked to choose one of the two options (A: left, B: right) themselves, followed by a second video and a second route choice. Following these, they were asked two final questions: why they chose the path they did, and whether they noticed a difference between the two videos. We also recorded the time that it took to complete each section. The exact layout and wording of the questionnaire can be found in the Appendix.

\begin{figure}
    \centering
    \includegraphics[width=0.5\linewidth]{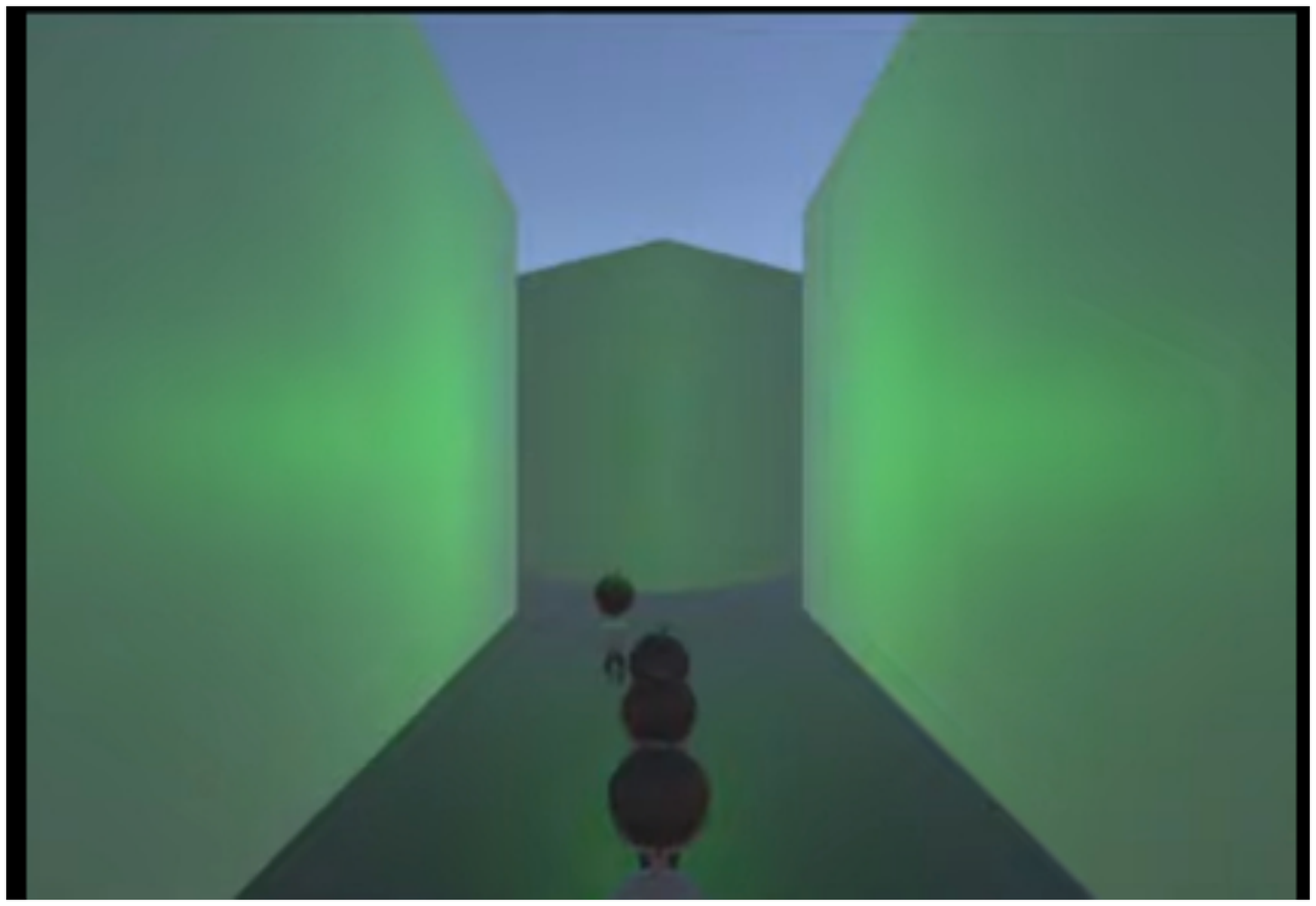}
    \caption{Screenshot of the virtual environment as seen by the participants.}
    \label{fig:VR_sample}
\end{figure}

The videos shown to participants were recorded using Unity 3D (Version 2019.3), and the virtual pedestrians were animated using the Unity Character Pack (sample Sam, Version 2.0.0) \cite{manual2017url}. Each video shows a corridor with high walls, that separates into two different unmarked exits at its end and with queuing computer-generated pedestrians, as shown in Figure \ref{fig:VR_sample}. In the videos eleven computer-generated pedestrians queued in the corridor and each chose one of the paths in a pre-defined sequence (see below). Flashing lights were added to create a state of urgency. All computer-generated pedestrians were only viewed from behind, they looked identical, wore a t-shirt and trousers, and had short hair. Ethical approval for our experiment was granted by the Research Ethics Committee of the Faculty of Engineering and Physical Sciences at the University of Leeds.

\subsubsection{Experimental scenarios}

To study the influence of social information on decision-making under uncertainty and to determine whether people pay more attention to the most recent decisions or to the majority of prior decisions when making their own decision, we showed different sequences of decisions by computer-generated pedestrians to participants.

The participants saw one of four different scenarios in the first video, that differed in the last two observed decisions by the eleven computer-generated pedestrians. The nine first decisions were identical and suggested a slight preference for route A: ABBABAABA. The last two decisions differed as follows:
\begin{enumerate}
    \item AAA: majority on route A, and route A the most recent previous decision.
    \item ABA: majority on route A, route A the most recent previous decision, but B before that.
    \item AAB: majority on route A, and route B the most recent previous decision.
    \item ABB: majority on route B, and route B the most recent previous decision.
\end{enumerate}

After the video, the participants were asked (in a separate page) to choose between the two options themselves. Following this, they were shown a second video; this was done to investigate potential learning or habituation effects. The last decisions differed as follows in their last three choices (the first nine choices remained the same, i.e. ABBABAAB):
\begin{itemize}
    \item AAA followed by AAB
    \item AAB followed by ABA
    \item ABA followed by AAB
    \item ABB followed by AAB
\end{itemize}

Participants were assigned to the four scenarios in equal numbers. This experiment with controlled sequences and a substantial sample size for each scenario (see below) was designed to distinguish between the decision-making strategies defined above, whilst accounting for confounding factors, such as sex, and to additionally query the match between decision strategies indicated by choices after the videos, and in reported by participants in their free text answers.

\subsubsection{Data collection and analysis}
Participants were recruited using the platform Prolific \footnote{www.prolific.com; accessed in March 2025} in March 2025. They received a small monetary reward based on the UK national living wage at the time and an expected completion time of 5 minutes (\pounds11.44*5/60=\pounds1.00). Of the 400 participants that took part in the experiment, 18 did not provide their sex or did not identify as male or female. These participants were excluded, as their number is too low to facilitate robust statistical analysis in this study.

The data were downloaded from Prolific in the form of csv files and they can be found in the supplementary material. All data analysis was performed using the R programming environment, version 4.3.1 \cite{rprogramming}. Throughout, we adopt a significance threshold of $\alpha = 0.05$ in our discussion. However, since this is an arbitrary choice, we report all p-values.

\section{Results}
\subsection{Influence of the majority in decision making}

Out of the 382 included entries, a total of 196 provided a realistic number for their age which were overwhelmingly between 18 and 34 years with a median of 27 years and a few outliers over 50 years (see Figure \ref{fig:age_box} in the Appendix). Given this skew in the distribution and the lack of representative numbers of older participants, we suggest that age cannot be used as a predictor in decision-making. We also did not consider completion times in our analysis as these demonstrate no clear trend (see Figure \ref{fig:times} in the Appendix), however we provide all data to do so in the future.

\begin{figure}[H]
    \centering
    \includegraphics[width=0.5\linewidth]{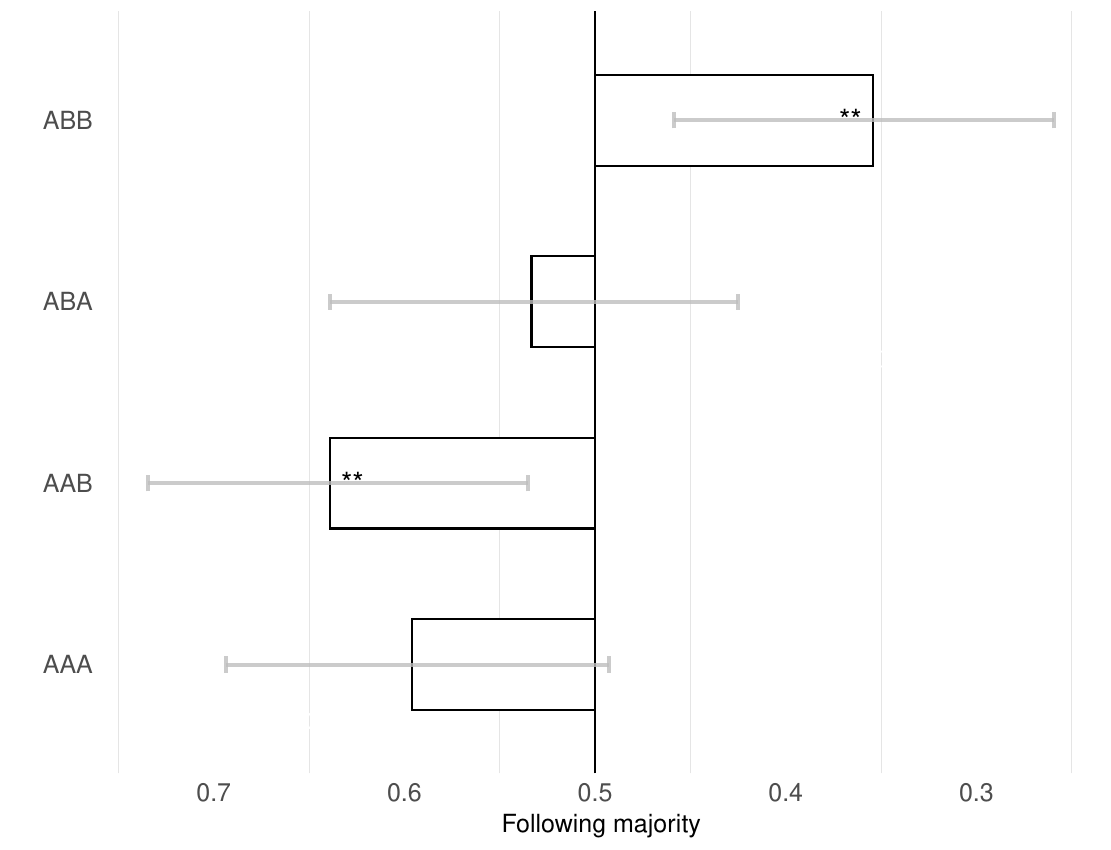}
    \caption{The experimentally determined probability of choosing route A in the four scenarios, centred at $0.5$. The asterisks at the end of the bars indicate a value of p-value $<0.01$ from a binomial test for the null hypothesis that $P(A)=0.5$. Error bars show 95\% confidence intervals. The number of participants for each scenario were: 99 (AAA); 97 (AAB); 90 (ABA); 96 (ABB).}
    \label{fig:bplot1}
\end{figure}

We expected the scenario to affect decision making, as the scenarios were designed to present social information via the majority and most recent decision that favoured options A and B alternately (recall section 2.2.1). Figure \ref{fig:bplot1} shows the proportions of participants choosing option A in our experiment, depending on the scenario. We notice that all scenarios with an overall majority on A result in $P(A)>0.5$, while the scenario with an overall majority on B instead has $P(A)<0.5$. While this demonstrates a connection between the overall majority and the final choice, we additionally notice that the scenarios where the most recent decision is A are not statistically different from a random choice (unlike the scenarios where the most recent decision is B are).

We used logistic regression to determine the effects of sex and scenario on route choices. Sex and scenario were expressed as categorical explanatory variables with two and four levels, respectively. We first used a Likelihood Ratio Test to establish that the sex variable and its interaction terms with scenario should be included in a model ($\chi^2_4=11, p=0.0265$). The results of logistic regression on the probability of choosing exit A for the explanatory variables scenario and sex, and their interaction terms are shown in Table~\ref{tab:modfull}.

We found that the changes of scenario ABA relative to the baseline (AAA for females) were significant, and reduced P(A). In scenario ABA, this effect was reversed for males (consider the sign of the sum of the effects `Scenario ABA', `Male', and `ScenarioABA:Male'). This sex-specific difference in behaviour may explain the findings for scenario ABA in figure~\ref{fig:bplot1}. An  explanation could be the possibly more male appearance of the computer-generated pedestrians. In scenarios AAA and ABA the last computer-generated pedestrian chooses option A, and the overall majority for A is $\Delta n = 3$ and $\Delta n = 1$, respectively. It is possible that in case of a less clear majority, women may prefer to not immediately follow a man. Even though previous studies have reported behavioural differences between sexes in simple route choice experiments \cite{bode2013human,bode2015increased}, this explanation remains speculation until further evidence becomes available. An alternative explanation could be the uncertainty that is being conveyed by the frequent switches; switches are already frequent in all the sequences, however in sequences AAA, AAB and ABB there seems to be a temporary convergence towards an option which does not happen in sequence ABA. It is possible that this is interpreted as a lack of trust in the decisions of others, that the participant reproduces in turn (for more on this, see Tables \ref{tab:probsa01}, \ref{tab:probsa05} and \ref{tab:probsa09} in the Appendix).

All other parameter-specific tests yielded non-significant p-values. These p-values can only be considered within the context of the model fitted. For example, the p-value for the Intercept in table~\ref{tab:modfull} should not be interpreted as contradicting the test on $P(A)$ in figure~\ref{fig:bplot1}, as different hypotheses are tested.

\begin{table}[h]
    \centering
       \caption{Logistic regression on route choice data with the response variable being a Boolean indicating whether option A was chosen. The intercept represents scenario AAA for females. The additional explanatory variables are shown in the table, they include scenario, sex, and interactions between scenario and sex (as indicated by colons, i.e. `ScenarioAAB:Male' indicates the effect of males in AAB compared to the baseline). The effect estimate, standard error, and results from Wald tests for the null hypothesis that the estimate is equal to zero are shown.  P-values lower than 0.05 are shown in bold face. Positive estimates imply increases in choosing option A, and vice-versa. For example, Scenario ABA led to a reduction in choosing A and this effect was unlikely to have arisen by chance (p=0.02287).}
    \begin{tabular}{lcccc}
        \hline
        & Estimate & Std. Error & z value & Pr($>|z|$) \\
        \hline
        (Intercept) & 0.42121 & 0.28084 &  1.500 & 0.13365 \\
        ScenarioAAB & 0.11287 & 0.41494 &  0.272 & 0.78561 \\
        ScenarioABA & -0.93204 & 0.40958 & -2.276 & \textbf{0.02287}\textsuperscript{*} \\
        ScenarioABB & -1.11436 & 0.38768 & -2.874 & \textbf{0.00405}\textsuperscript{**} \\
        Male & -0.06982 & 0.41053 & -0.170 & 0.86496 \\
        ScenarioAAB:Male & 0.14187 & 0.58966 &  0.241 & 0.80987 \\
        ScenarioABA:Male & 1.49693 & 0.61163 &  2.447 & \textbf{0.01439}\textsuperscript{*} \\
        ScenarioABB:Male & 0.33218 & 0.60571 &  0.548 & 0.58341 \\
        \hline
    \end{tabular}
 
    \label{tab:modfull}
\end{table}

\subsection{Self-reported decision strategies}

After choosing routes, we asked participants to explain how they made their choice. The purpose of this question was to gain a perspective on the decision-making strategies and to examine the match between participants' self-reported decision mechanisms and the evidence in our choice experiment. The only other work that examines this type of setting, and aims to answer the same question (which form of social information are decision makers using) has been performed with fish \cite{Kadak2020}, where such a comparison is not possible. We allowed free-form answers, which we classified into ten categories: balance, bias, first decision, majority, minority, most recent, pattern, random, unclear and N/A (missing answers). The categorisation was additionally evaluated using a LLM (Claude 3 Sonnet), and was confirmed as being sound with good category definitions especially in the cases representing majority, minority and most recent decisions. 

\begin{figure}[H]
    \centering
    \includegraphics[width=0.9\linewidth]{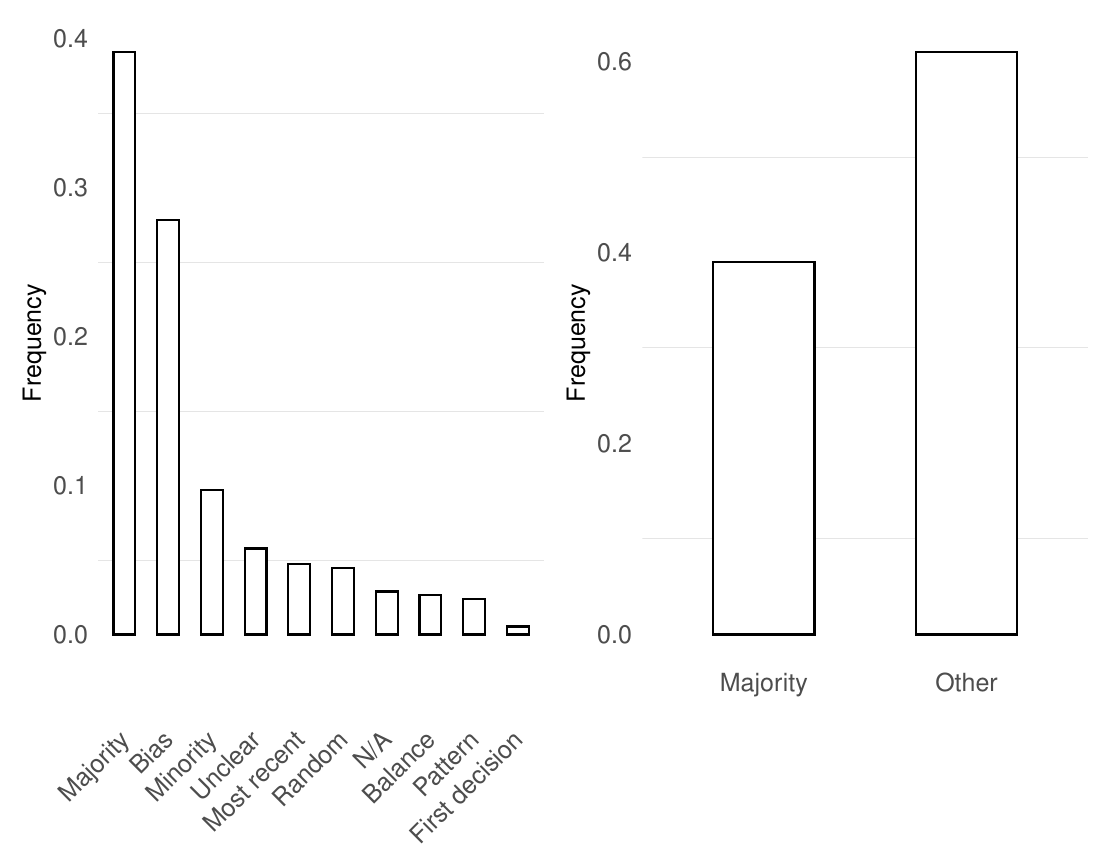}
    \caption{Distribution of self-reported strategies in route choice based on a manual classification (a), and simplified classification of two categories used in the analysts (b). See text for a more detailed description of the strategy classes.}
    \label{fig:strategies}
\end{figure}

While people can be poor at explaining their own motivations (as demonstrated especially in the `bias' category), we were able to establish a statistical connection between what people said and what they did. In agreement with our findings from the experiment (Figure~\ref{fig:bplot1}), we found that most people self-reportedly followed the majority in previous decisions (Figure \ref{fig:strategies}). Additionally, we observed that following the most recent decision was an extremely rare strategy, with some participants actually reporting actively contradicting it. Any other self-reported strategy was equally unlikely compared to following the majority, apart from those choosing based on bias. However, we note that what we have labelled as `bias' here is a variety of different answers; some referring to `intuition', some to their dominant hand, some to a `gut feeling' etc, preventing cohesion of these answers and as a consequence an evaluation of them.

To formally investigate the correlation between self-reported decisions based on the majority, we extended our statistical analysis by performing binomial tests to establish whether there is a significant difference in how often those who claim to follow the majority actually the majoritarian option in their scenario (either A in scenarios AAA, AAB and ABA or B in scenario ABB) with a probability significantly different to $0.5$, versus how often those following other strategies do. Out of 149 participants reporting following the majority, 116 successfully did so; out of the 233 participants that reported using other decision strategies, 115 ended up choosing the majoritarian option in their scenario. The binomial tests showed that those who claim to follow the majority actually do so with a probability of 0.779, with the binomial test showing a p-value of 5.018e-12. On the other hand, those claiming to employ different strategies follow the majority with a probability of 0.494 with the binomial test showing a p-value of 0.8958. A further two-sided proportion test between the two groups (those self-reporting following the majority, and the rest) reveals that following the majority leads to a statistically significant rate of choosing the majoritarian option, compared to the other strategies as the proportion test produces a p-value of 5.062e-08.

\subsection{Habituation or learning effects}
The purpose of presenting a second video to participants was to investigate habituation or learning effects. We measured whether the probability of choosing option A when observing the sequence AAB in the second video was affected by the scenario the participants had seen before. We found no indication of such effects (Table \ref{tab:AAB}).

\begin{table}[h]
    \centering  
    \caption{Logistic regression on route choice data from the decision after viewing the second video in the experiment. We only use data where the second scenario viewed was AAB ($n=285$; i.e. we exclude data where the first scenario was AAB). The intercept now affects the probabilities of females to choose A in the second video given that the first video was scenario AAA. For further details on how to interpret the table see Table~\ref{tab:modfull}. Large p-values suggest prior experience of different scenarios did not affect route choice in our experiment.}
    \begin{tabular}{lrrrr}
        \hline
        & Estimate & Std. Error & z value & Pr($>|z|$) \\
        \hline
        (Intercept) &  0.18924 & 0.27595 &  0.686 & 0.493 \\
        ScenarioABA & 0.06207 & 0.40100 &  0.155 & 0.877 \\
        ScenarioABB & -0.15749 & 0.37371 & -0.421 & 0.673 \\
        Male & -0.27625 & 0.40407 & -0.684 & 0.494 \\
        ScenarioABA:Male & 0.12025 & 0.58599 &  0.205 & 0.837 \\
        ScenarioABB:Male & 0.18388 & 0.59000 &  0.312 & 0.755 \\
        \hline
    \end{tabular}

    \label{tab:AAB}
\end{table}

\subsection{Empirical evidence for decision-making strategies}

Rewriting the decision-making model from Equation \ref{model_basic} as follows:
\begin{equation}
    \frac{1}{1+e^{\log(a) + \log(s^{I})}}
\end{equation} 
and recalling that $I$ captures the social information (i.e., $\Delta n$, $\textrm{sign}(\Delta n)$, $d$), we can see that the decision-making models introduced above can be fitted via established Maximum Likelihood methods for logistic regression. Thus, we fit the models for the aggregate, majority, and dynamic decision-making strategies to our data, estimating the parameters $a,s$ for each of the strategies, and we use the Akaike Information Criterion (AIC) to determine which strategy fits our data best. Note that the decision-making models that we consider have the same number of parameters, and so the model with the lowest AIC will coincide with the highest likelihood.

 \begin{figure}[H]
    \centering
    \includegraphics[width=\linewidth]{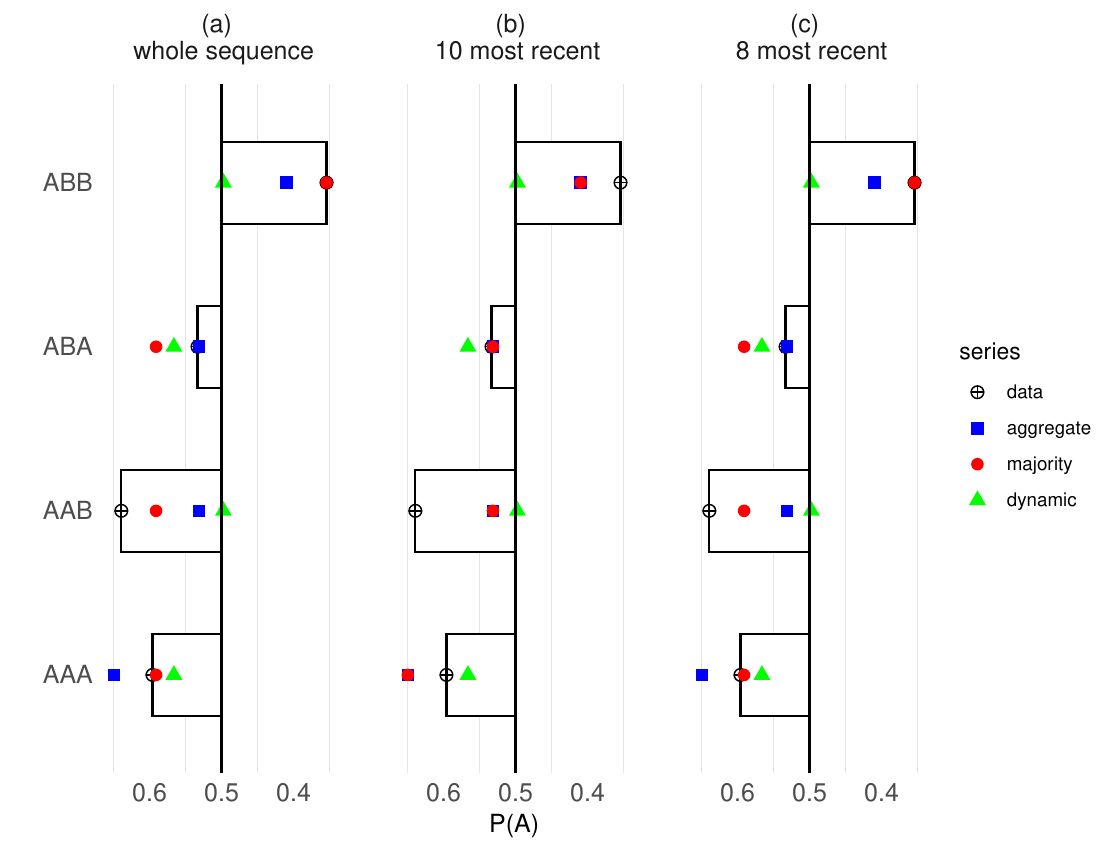}
    \caption{The proportion of participants choosing route A, P(A), for the different scenarios. We show P(A) for our experimental data (compare to Figure \ref{fig:bplot1}) and the fitted values for the three different theoretical models for decision-making strategies introduced in the text. The offset along the x-axis is for clarity of illustration.}
    \label{fig:modelfit}
    \end{figure}
    
We perform this fit for different assumptions of the amount of prior decisions the focal individual is able to keep track of (ranging from only the three most recent, to the complete sequence). Figure \ref{fig:modelfit} shows that the different decision-making strategy models result in a fit of varying accuracy to the data for all these cases; panel (a) shows the case where the participant can observe the whole sequence, which also coincides with the case where she can observe the 9, 7, 5 and 3 most recent decisions since the value of $\Delta n$ is the same for all these cases. Panel (b) shows the case where the participant can observe the 10 most recent decisions, which also coincides with observing the 4 most recent. Lastly, panel (c) is the case where the participant can observe the 8 most recent decisions, which also coincides with the case where she can observe the 6 most recent ones. The dynamic strategy appears to deviate most from the data for at least the two scenarios where the last decision in sequence was for option B (AAB and ABB), suggesting participants are unlikely to show this behaviour. This is also confirmed by this model having the highest AIC value out of the three strategies (table~\ref{tab:modelfit}). The aggregate and majority strategies provide a reasonable fit for all scenarios but comparing AICs suggests that the majority strategy provides the best fit (see Tables~\ref{tab:modelfit}, \ref{tab:modelfit_last10}, \ref{tab:modelfit_last8}). The majority strategy is static, and unaware of the sequence of decisions. Our finding contrasts with the results of Kadak and Miller \cite{Kadak2020}, as well as some theoretical predictions [\cite{mann2014humbugs, mann2021}] that expect the most recent decisions to be more influential than the majority decision where these conflict.

\begin{table}[h]
    \centering
       \caption{Fitting decision-strategy models to experimental data, for the assumption that the participants can observe the full sequence (equivalent to observing most recent 9, 7, 5 or 3 decisions). We show the parameter estimates and Akaike Information Criterion for each model based on Maximum Likelihood Estimation.}
    \begin{tabular}{lccc}
        \hline
        Model & $a$ value & $s$ value & AIC \\
        \hline
        Majority & 0.89 & 1.623 & 515.77 \\
        Aggregate & 0.887 & 1.279 & 520.66 \\
        Dynamic & 1.136 & 1.149 & 530.24 \\
        \hline
    \end{tabular}
 
    \label{tab:modelfit}
\end{table}

\begin{table}[h]
    \centering
        \caption{Fitting decision-strategy models to experimental data for the assumption that the participants can observe the 10 more recent decisions (equivalent to observing the 4 more recent decisions).}
    \begin{tabular}{lccc}
        \hline
        Model & $a$ value & $s$ value & AIC \\
        \hline
        Majority & 1.134 & 1.636 & 520.66 \\
        Aggregate & 1.134 & 1.279 & 520.66 \\
        Dynamic & 1.136 & 1.149 & 530.24 \\
        \hline
    \end{tabular}

    \label{tab:modelfit_last10}
\end{table}

\begin{table}[h]
    \centering
        \caption{Fitting decision-strategy models to experimental data for the assumption that the participants can observe the 8 more recent decisions (equivalent to observing the 6 more recent decisions). }
    \begin{tabular}{lccc}
        \hline
        Model & $a$ value & $s$ value & AIC \\
        \hline
        Majority & 0.548 & 2.634 & 515.77 \\
        Aggregate & 0.694 & 1.279 & 520.66 \\
        Dynamic & 1.136 & 1.148 & 530.24 \\
        \hline
    \end{tabular}

    \label{tab:modelfit_last8}
\end{table}

These values deviate from the ones in \cite{Kadak2020}, where they had assumed that $a=1$ prior to the fitting due to the identical nature of the two options, resulting in the existence of only one free parameter referring to the strength of social following (here represented by $s$). Here, by allowing for the environmental reliability $a$ to be a free parameter, we instead get a pair of optimal values for each decision rule. We consider that relaxing the assumption that $a=1$ to be reasonable, since many of the participants reported having an inherent sense of bias despite the environment being completely symmetrical. This way the definition of $a$ remains in line with its original definition, where it is equal to the ratio of choosing option B given the non-social information, over choosing option A given the non-social information. However, we also repeated the fit with the assumption of $a=1$ for a direct comparison to the work of Kadak and Miller. Again, we find that the majority model is the best fit to the data (see Table~\ref{tab:modelfit_2} in the Appendix).
%, and overall the AIC is higher for all models compared to the case where there is no prior assumption for $a$ indicating that the assumption $a=1$ does not lead to the best fit.

The estimates for the model parameters in both cases are in line with the constraints mentioned previously, where $s>1$ and $a > 0$. In all cases the intensity of social following is higher for the majority strategy and lower for the dynamic strategy. Note that in all cases $a \neq 1$, indicating an inherent bias of the participants to choose option B instead. We also note that for the dynamic strategy (and all strategies in the case where the last 10 or 8 decisions are observed, \ref{tab:modelfit_last10}) the best fit is for $a>1$, which corresponds to a slight bias towards B.

\section{Discussion}
We conducted an online experiment, where participants were shown a virtual evacuation scenario where eleven decisions by others were shown. They were then asked to make their own choice. By collecting data from four different treatments, we were able to identify key patterns related to the mechanisms of use of social information in sequential decision making.

The underlying mechanisms of sequential decision-making have been the focus of several theoretical publications, however there are very limited experimental works on it. To our knowledge, only Kadak and Miller  \cite{Kadak2020} have focused specifically on this model set-up and this question. In this research they suggest that dynamic strategies, such as following the most recent decision, are more plausible than static ones. We found that both our data and the answers provided by our participants suggest the use of static information instead, namely the overall majority. We found this to be the case, regardless of the fraction of the prior sequence they are able to observe, and regardless of whether we assume $a=1$. While we found some evidence of dynamic information being used, possibly related to gender, following the majority remained the dominant mechanism.

The contradiction to the work of Kadak and Miller  \cite{Kadak2020} can be due to a number of reasons: 1) The experimental subjects belong in different species, with significantly different capacities so it's possible that humans are simply more capable of keeping track of longer sequences and hence strategies that notice more than one prior decision are more accessible. 2) The sequences we investigated were not identical to those of \cite{Kadak2020} , with ours being longer and more likely to include apparent weak social following by those ahead of the focal individual (i.e. many switches between A and B). This may have conveyed a greater uncertainty to the focal observer about whether following the agent directly in front was a good strategy, compared to the case where she would see a shorter sequence with fewer switches; note that the sequence conveying a similar type of indecisiveness in Kadak and Miller \cite{Kadak2020} is also associated with a significantly lower probability of the fish choosing option A (specifically, their sequence ABAB leads to a lower $P(A)$ than sequence AABB). Sequences with frequent switches may indicate a weak social response that does not depend strongly on the order of decisions \cite{Mann10388}, and the participants observing them can infer, and in turn reproduce, this weak social response. A wider range of different sequences of decisions would be desirable to more comprehensively contrast the models decision-making strategies (specifically, a wider range of $\Delta n$ values would be useful). Our results also challenge the theoretical expectation that more recent decisions should carry greater weight \cite{Mann2018} --but agree with other work that finds that strategies relying on aggregated versions of the available sequential information lead to comparable results \cite{Perez-Escudero2011}. While further experimental work is necessary to provide a clearer picture, it is also likely that there is not one single strategy that vertebrate use, but that the strategy depends on key factors such as computational capacities, memory, etc.

One of the main obstacles working with animals is that we do not know their internal mechanisms and can thus just attempt to deduce them from observations, a process susceptible to uncertainty. However, working with humans presents the advantage of gaining insight into their internal mechanisms. The free-form answers we received revealed that many of the participants were able to identify patterns such as which is the majoritarian option, with a large fraction of them choosing to follow it. However, it also revealed that an equally large fraction provided reasons that referred to qualitative things, such as a ``gut feeling", ``intuition", or their dominant hand revealing that more than just the social information factors in decision-making under uncertainty. This makes the choice to leave $a$ as a free parameter reasonable, as even though the environment is symmetric the participants' perception of it isn't; this also probably means that the value of $a$ will not be shared between participants; it might follow a distribution, with the normal distribution being an intuitive starting point for future research. Additionally, while ``bias" (as vague as it may be) proves important, we find that the bias is a pre-existing condition for the participants instead of newly acquired, as our data didn't reveal any habituation or learning effects. Lastly, the variety of answers reveals that there exists more than one strategy that humans follow in this type of setting; for instance, some of the participants explicitly choosing to go to the minority option as it may be less congested. This concept could be explored further by adopting it to models for crowd behaviour.

Despite these insights, several limitations of the present study should be acknowledged. While virtual experiments are a well-accepted experimental paradigm due to their benefits in terms of ease of data collection, safety, and controllability, at the same time they have some inherent limitations. They are highly abstracted, leading to a non-realistic environment for the participants that lacks elementary aspects that would otherwise affect behaviour (such as visual or auditory stimuli or a sense of urgency). Additionally, the homogeneous appearance of the virtual pedestrians are taking away complexity, since in a real crowd the people would vary in age, size, speed, and appearance. Such an environment may lead to different results, although it is worth noting that even in this virtual and simplified setting we were able to elicit behavioural differences, possibly including gender effects. Moreover, our experiments did not incorporate dynamic elements such as time pressure, congestion, and  perceived urgency, although some participants mentioned these factors in their qualitative responses. These aspects are likely to influence decision-making and should be included in future work to assess how time constraints and physical interactions alter participants' use of social information during evacuations. Additionally, the choice of treatments in our design was intentionally limited to a small set of carefully constructed decision sequences, which allowed us to distinguish between different behavioural mechanisms and to control for potential confounds such as gender. However, a wider range of sequences, including a broader distribution of $\Delta n$ values, would be valuable for a more comprehensive comparison of theoretical models and for capturing potential complicated responses to social information. Finally, the participant pool consisted primarily of young adults, which may limit the generalisability of our findings to broader populations as previous studies have established the age differences in pedestrian navigational skills and performance \cite{xu2024age}.

\section{Conclusion}
This study provides empirical evidence on how pedestrians use social information after observing the sequential route choices of others. Using a virtual experiment with 382 participants, we found that, unlike previous theoretical predictions and animal studies, humans predominantly follow the majority of observed decisions rather than placing greater weight on the most recent ones. These findings highlight a distinct social mechanism underlying human social decisions, suggesting differences in how humans integrate social cues compared to animals. Moreover, gender differences emerged in how individuals reacted to the social information, indicating potential variability in sensitivity to social information. Future research should extend this approach to more ecologically valid settings and incorporate these behavioural mechanisms into predictive models of collective evacuation and crowd dynamics.

\section{Acknowledgments}
This work was supported by UK Research and Innovation Future Leaders Fellowship (MR/S032525/1 and MR/X036863/1).

\section{Competing Interests}
The authors declare no competing interests.

\section{Ethical Approval}
Ethical approval was obtained by the Research Ethics Committee of the Faculty of Engineering and Physical Sciences of the University of Leeds on 22/3/2024, and has the reference ``ID LTMATH-003''.

\appendix
\section{Appendix} 

\renewcommand\thetable{\Alph{section}.\arabic{table}}   
\renewcommand\thefigure{\Alph{section}.\arabic{figure}} 
\setcounter{figure}{0} 
\setcounter{table}{0} 

\subsection{Age distribution}
\begin{figure}[H]
    \centering
    \includegraphics[width=0.5\linewidth]{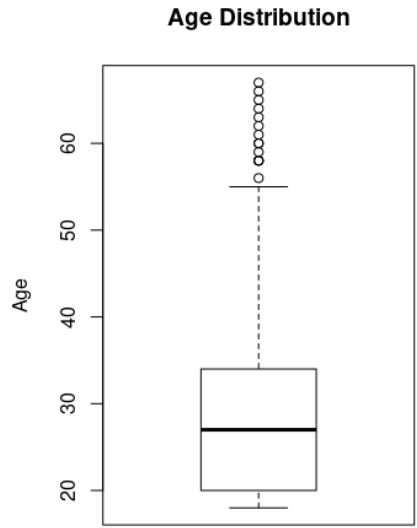}
    \caption{Age distribution of participants. The median age was 27 years all, with most participants falling into the range between 18-34 and a few outliers over 50. Consequently the age of the participants is not informative enough to be considered as a factor in decision-making in this context.}
    \label{fig:age_box}
\end{figure}

As part of our survey we collected the participants' age, the time they took to watch the video (note, the video was prerecorded, but to move to the next part of the survey the participants needed to press a button) and the required to decide on an option after watching the video. Figure~\ref{fig:age_box} shows the participants' age distribution. It shows that this followed a skewed normal distribution with a median age of 27 and a small number of outliers of ages older than 50. From this, we conclude that the age range does not provide useful information for this experiment.

\subsection{Completion times}
\begin{figure}[H]
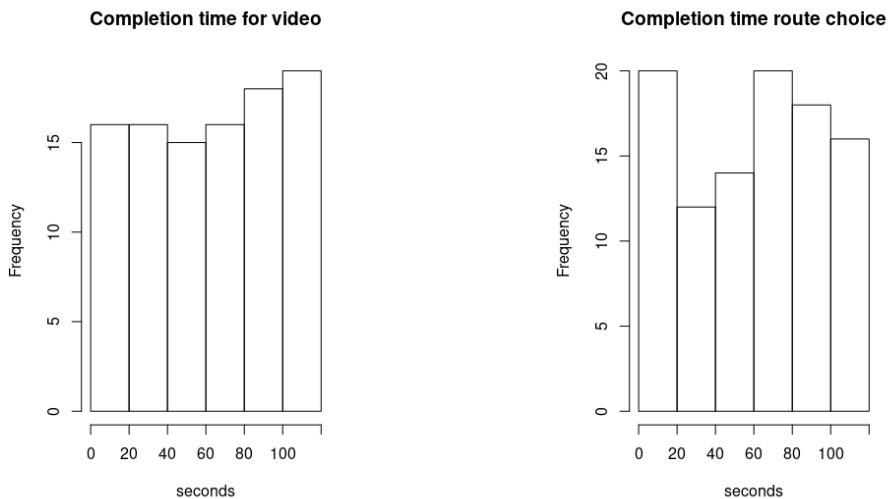

    \centering
    \begin{subfigure}[b]{0.4\textwidth}
        \includegraphics[width=\textwidth]{plots/video_timeHist.pdf}
        %\caption{Time to watch video (in seconds)}
        \end{subfigure}
    \hfill
    \begin{subfigure}[b]{0.4\textwidth}
        \includegraphics[width=\textwidth]{plots/Q7_timeHist.pdf}
        %\caption{Time to choose route (in seconds)}
    \end{subfigure}
    \caption{Time (in seconds) required for watching video \#1 (panel a) and for submitting the path choice (panel b). The time required for both shows no clear trend.}
    \label{fig:times}
\end{figure}

Figure~\ref{fig:times} show the time (in seconds) required to watch the video and make a path decision. It's clear that these times show no clear trend, and in addition to the uncertainty of the reason behind these durations, this data is not informative for the purpose of this study.

\subsection{Model fit for $a=1$}

Table \ref{tab:modelfit_2} summarises the optimal values of the parameter $s$ and AIC values for each strategy, for the case where the participant is assumed to be observing the full sequence and for $a=1$.

\begin{table}[H]
  \centering
    \caption{Fitting decision-strategy models to experimental data, with the assumption that a=1. We show the parameter estimates and Akaike Information Criterion for each model based on Maximum Likelihood Estimation.}
  \renewcommand{\arraystretch}{1.2}
  \begin{tabular}{|p{2cm}|c|c|c|c|c|c|c|}
    \hline
    \multirow{2}{5cm}{\textbf{Model}} & \multicolumn{2}{c|}{\textbf{full seq}} & \textbf{last 10} & \textbf{last 8}\\
    % \hline
    % \textbf{Inactive Modes} & \textbf{Description}\\
    \cline{2-7}
    & \textbf{s value} & \textbf{AIC} & \textbf{s value} & \textbf{AIC} & \textbf{s value} & \textbf{AIC}\\
    %\hhline{~--}
    \hline
    Majority & 1.53 & 514.69 & 1.635 & 520.12 & 1.444 & 522.06 \\ \hline
    Aggregate & 1.229 & 519.56 & 1.279 & 520.12 & 1.132 & 522.85 \\ \hline
    Dynamic & 1.146 & 529.79 & 1.146 & 529.79 & 1.146 & 529.79 \\ \hline
  \end{tabular}

  \label{tab:modelfit_2}
\end{table}

\subsection{Effect of ordered decisions}

Here, we work out the probabilities of the sequences we chose. While we expect the use of the dependencies strategy to be unlikely, since simplified versions offer good approximations [\cite{Perez-Escudero2011}], these sequences are visible to the participants, and may convey information to the observed about the degree of trust towards the decisions of others; for instance, if there is little indication of social following by the participant, she may in turn not follow the social information as the cues available to her indicate against it. 

The dependencies strategy has the following formulation:
\begin{equation}
    P_A = \frac{1}{1+aS}
    \label{model_basic2}
\end{equation}
where $S$ is a product of all the prior decisions, but with parameter $\tilde{a}$ instead of $a$. This parameter represents the assumption that the focal individual makes about the other agents' personal information; if $\tilde{a} < a$ she assumes that others have better personal information than her and follows them strongly, and vice versa (for more details on this strategy, see \cite{Perez-Escudero2011}).

For 11 prior decisions, there are $2^{11}$ possible sequences one can observe; the sequences considered here are: AAAAAAAAAAA (all A), BBBBBBBBBBB (all B), ABBABAABAAA (AA), ABBABAABAAB (AB), ABBABAABABA (BA) and ABBABAABABB (BB). To calculate these, we used the code provided in the supplementary material of \cite{Perez-Escudero2011} as re-written by \cite{sigalou2023evolutionary}. Depending on the values of the parameters $a, \tilde{a}$ each sequence has different probabilities of occurring. For example, for low values of $a$ and $\tilde{a}$ (i.e., for reliable environmental information and strong social following), sequences that include long strings of option A are extremely probable while sequences that involve frequent switches or many B choices are improbable; for high $a$ and $\tilde{a}$ (i.e. for unreliable environmental information and weak social following), sequences with long strings of option A become more improbable while sequences that involve frequent switches become more probable (See Tables \ref{tab:probsa01}, \ref{tab:probsa05} and \ref{tab:probsa09}).  
These probabilities provide relevant information about our experiments; 

\begin{table}[H]
    \centering
    \begin{tabular}{lccc}
        \hline
        Sequence & $\tilde{a}=0.1$ & $\tilde{a}=0.5$ & $\tilde{a}=0.9$\\
        \hline
        all A & 0.8846 & 0.7729 & 0.5417 \\
        all B & 2.68e-06 & 0.0181 & 0.2873 \\
        AA & 1e-07 & 2e-05 & 4e-05 \\
        AB & 5e-08 & 3.36e-06 & 1e-05 \\
        BA & 2.91282e-09 & 2.21e-06 & 1e-05 \\
        BB & 5.644e-11 & 3e-07 & 8.5e-07 \\
        \hline
    \end{tabular}
    \caption{Probabilities of sequences using the dependencies model, for $a=0.1$}
    \label{tab:probsa01}
\end{table}

\begin{table}[H]
    \centering
    \begin{tabular}{lccc}
        \hline
        Sequence & $\tilde{a}=0.1$ & $\tilde{a}=0.5$ & $\tilde{a}=0.9$\\
        \hline
        all A & 0.582476 & 0.308622 & 0.0669 \\
        all B & 2.7334e-11 & 1e-05 & 0.0069 \\
        AA & 3.9e-07 & 0.000102 & 0.00043 \\
        AB & 1e-06 & 0.000104 & 0.00034 \\
        BA & 1.2e-07 & 7.8e-05 & 0.00032 \\
        BB & 1e-08 & 5.3e-05 & 0.00024 \\
        \hline
    \end{tabular}
    \caption{Probabilities of sequences using the dependencies model, for $a=0.5$}
    \label{tab:probsa05}
\end{table}

\begin{table}[H]
    \centering
    \begin{tabular}{lccc}
        \hline
        Sequence & $\tilde{a}=0.1$ & $\tilde{a}=0.5$ & $\tilde{a}=0.9$\\
        \hline
        all A & 0.4138 & 0.1402 & 0.0124 \\
        all B & 2.45e-13 & 2e-07 & 0.00051 \\
        AA & 3e-07 & 5e-05 & 0.00025 \\
        AB & 1.35e-06 & 9.219e-05 & 0.00035 \\
        BA & 2.419e-07 & 7.285e-05 & 0.00032 \\
        BB & 4.22e-08 & 8.992e-05 & 0.00045 \\
        \hline
    \end{tabular}
    \caption{Probabilities of sequences using the dependencies model, for $a=0.9$}
    \label{tab:probsa09}
\end{table}

It's clear that the relative probability of any sequence depends on the values of $a, \tilde{a}$. Overall, it decreases as $a, \tilde{a}$ increases; that is, it becomes more probable to observe more switches in uncertain environments and in environments with less weak social following. Here, for all four scenarios the virtual environment is symmetrical, which corresponds to $a=1$ in the model. In that case, these sequences are more likely for values of $\tilde{a}$ closer to 1, which indicates less trust in others' decisions. Intuitively, if the participants observe others failing to follow each other, they also become skeptical of doing so themselves.

\newpage
\printbibliography
\end{document}